\documentclass{article}

\usepackage{amsmath, amsthm, amssymb}
\usepackage{graphicx}
\usepackage{graphics}
\usepackage{subfig}

\newtheorem{theorem}{Theorem}[section]

\newtheorem{proposition}[theorem]{Proposition}

\begin{document}

\title{Quantum mechanical inverse scattering problem at fixed energy:
a constructive method}

\author{Tam\'as P\'almai\footnote{Electronic mail:
palmai@phy.bme.hu} {\ }and Barnab\'as Apagyi\footnote{Electronic mail:
apagyi@phy.bme.hu} \\ \ \\
Department of Theoretical Physics\\
Budapest University of Technology and Economics\\
Budafoki ut 8., H-1111 Budapest, Hungary}

\maketitle

\begin{abstract}The inverse scattering problem of the three-dimensional Schr\"odinger equation is considered at fixed scattering energy with spherically symmetric potentials. The phase shifts determine the potential therefore a constructive scheme for recovering the scattering potential from a {\it finite} set of phase shifts at a fixed energy is of interest. Such a scheme is suggested by Cox and Thompson \cite{CT1} and their method is revisited here. Also some new results are added arising from investigation of asymptotics of potentials and concerning statistics of colliding particles. A condition is given \cite{PAJMP} for the construction of  potentials belonging to the class $L_{1,1}$ which are the physically meaningful ones. An uniqueness theorem is obtained \cite{PAJMP} in the special case of one given phase shift by applying the previous condition. It is shown that if only one phase shift is specified for the inversion procedure the unique potential obtained by the Cox-Thompson scheme yields the one specified phase shift while the others are small in a certain sense. The case of two given phase shifts is also discussed by numerical treatment and synthetic examples are given to illustrate the results. Besides the new results this contribution provides a systematic treatment of the CT method.
\end{abstract}

\newpage
\section{General results, introduction}

We start with the Schr\"odinger equation in $\mathbb{R}^3$ at a fixed positive
energy
\begin{equation}
[\nabla^2+1-q(x)]\Psi(x,\alpha)=0\qquad\text{in }\mathbb{R}^3
\end{equation}
whose scattering solution takes the form
\begin{equation}
\Psi(x,\alpha)=e^{i x\cdot\alpha}+A(\alpha',\alpha)\frac{e^{i r}}{r}+o\left(\frac{1}{r}\right),\quad r=|x|\to\infty,\quad \alpha'=\frac{x}{r},
\end{equation}
where $\alpha$ is the direction of the incident wave and $A(\alpha,\alpha')$
is the scattering amplitude. The following general theorem is due to Ramm \cite{RammIP}.
\begin{theorem}
$A(\alpha',\alpha)$ scattering amplitude $\forall \alpha'\in\tilde{S}_1^2$, $\forall \alpha\in\tilde{S}_2^2$ (arbitrary small open subsets of $S^2$) determine $q(x)$ uniquely in the function class $Q_a=\{q:q=\bar q,\,q(x),\,|x|>a,\,q(x)\in L^2(B_a)\}$, $B_a=\{x:x\in\mathbb{R}^3,\,|x|<a\}$.
\end{theorem}

The present treatment is restricted to spherically symmetric potentials, that
is $q(x)=q(r)$. In this case we have the partial wave expansion of the wave
function:
\begin{equation}
\Psi(x,\alpha)=\sum_{\ell=0}^\infty\sum_{m=-\ell}^\ell 4\pi i^\ell \frac{\psi_\ell(r)}{r}Y_{\ell m}(x/r)\bar{Y}_{\ell m}(\alpha),
\end{equation}
where $\psi_\ell(x)$ satisfies the radial Schr\"odinger equation (see below) with the appropriate boundary conditions.

For spherically symmetric potentials the scattering amplitude takes a similar
expansion form, namely
\begin{align*}
A(\alpha',\alpha)=A(\alpha'\cdot\alpha)=\sum_{\ell=0}^\infty\sum_{m=-\ell}^{\ell}A_\ell Y_{\ell m}(\alpha')\bar{Y}_{\ell m}(\alpha),\\
A_\ell=2\pi i(1-e^{2 i\delta_\ell})=4\pi e^{i\delta_\ell}\sin\delta_\ell.
\end{align*}
The phase shifts, $\{\delta_\ell\}_{\ell=0,1,2,\ldots}$ can be restricted to
$-\frac{\pi}{2}\leq\delta_\ell\leq\frac{\pi}{2}$, since from the experimental
point of view they are undetermined to an additive factor of
$k\pi$, $k\in\mathbb{Z}$. Then it is apparent that the knowledge of the phase
shifts is equivalent to that of the scattering amplitude. However not all the phase shifts (as considered in the theorem of Loeffel \cite{Loeffel}) are needed for unique reconstruction of the potential.
The following result is due to Horv\'ath and Ramm \cite{HorvTAMS,RammCMP}.
\begin{theorem}
The phase shifts $\{\delta_\ell\}_{\ell\in\mathfrak{L}}$ determine the
potential uniquely in the class
$Q=\{q:q=\bar{q},\,q(x)=q(r),\,r=|x|,\,q(r)=0,\,r>a;\,rq(r)\in L_{1}(0,a)\}$
if the M\"untz condition holds for $\mathfrak{L}$:
\begin{equation}
\sum_{\ell\in\mathfrak{L},\ell\neq0}\frac{1}{\ell}=\infty.
\end{equation}
This condition is almost necessary in the sense that in the class $Q_\sigma$
with $0<\sigma<2$ the set $\{\delta_\ell\}$ with $\sum \ell^{-1}<\infty$ is
not enough to recover the potential uniquely.
$Q_\sigma=\{q:q=\bar{q},\,q(x)=q(r),\,r=|x|,\,q(r)=0,\,r>a;\,r^{1-\sigma}q(r)\in L_{1}(0,a)\}$.
\end{theorem}
It is because of this result that one may look for a potential when a part of
the scattering phase shifts are known.

The phase shift $\delta_\ell$ appears in the asymptotic form of the regular
solution of the Schr\"odinger equation, which is defined in the following
manner:
\begin{align*}
L_r\varphi_\ell(r)=\ell(\ell+1)\varphi_\ell(r),\qquad L_r=\left[r^2\frac{d^2}{d r^2}+r^2-r^2q(r)\right],\\
\varphi_\ell(r)=\frac{r^{\ell+1}}{(2\ell+1)!!}+o(r^{\ell+1}),\qquad r\to0,\\
\varphi_\ell(r)=B_\ell\sin(r-\frac{\ell\pi}{2}+\delta_\ell)+o(1),\qquad r\to\infty
\end{align*}
with $B_\ell$ being a constant. Note that for the $q\equiv0$ zero potential
we have $L_{r0}u_\ell(r)=\ell(\ell+1)u_\ell(r)$, where
$L_{r0}=\left[r^2\frac{\partial^2}{\partial r^2}+r^2\right]$ and
$u_\ell(r)=\sqrt{\frac{\pi r}{2}}J_{\ell+\frac{1}{2}}(r)$ and $J_n(r)$
is the $n$th order Bessel function of the first kind.

The Povzner-Levitan representation in the fixed energy problem \cite{CS} has long been used, however the existence and uniqueness of a $K(r,r')\in C^2(\mathbb{R}^+\times\mathbb{R}^+)$ transformation kernel have been proven rigorously only recently \cite{RammCMP} for potential $q(r)\in C^1(0,a)$ independent of $\ell$ and satisfying
\begin{equation}\label{PL}
\varphi_\ell(r)=u_\ell(r)-\int_0^r K(r,\rho)u_\ell(\rho)\rho^{-2}d\rho,\qquad K(r,0)=0.
\end{equation}
For $K(r,r')$ we have a Goursat-type problem (equivalent to the Schr\"odinger
equation):
\begin{align*}
L_r K(r,r')=L_{r'0}K(r,r'),\qquad 0<r'\leq r,\\
q(r)=-\frac{2}{r}\frac{d}{dr}\frac{K(r,r)}{r},\qquad K(r,0)=0.
\end{align*}

Now, in analogy with the fixed--$\ell$ problem a
Gel'fand-Levitan-Marchenko-type integral equation is written up for $K(r,r')$:
\begin{equation}\label{GLM}
K(r,r')=g(r,r')-\int_0^r d\rho \rho^{-2}K(r,\rho)g(\rho,r'),\qquad r\geq r'.
\end{equation}
Since $K(r,r')$ is unique this integral equation must have a unique solution
for $K(r,r')$ to make sense. Furthermore, for consistency with the original
Schr\"odinger equation, $g(r,r')$ must satisfy
\begin{equation}
L_{r0}g(r,r')=L_{r'0}g(r,r'),\qquad
g(r,0)=g(0,r')=0
\end{equation}
and through the integral equation
\begin{equation}
q(r)=-\frac{2}{r}\frac{d}{dr}\frac{K(r,r)}{r}
\end{equation}
is maintained. $g(r,r')$ is a good candidate for approximation since its differential equation can be satisfied trivially. For instance if the angular momentum expansion \cite{CS}
\begin{equation}
g(r,r')=\sum_\ell c_\ell \gamma_\ell(r,r'),
\end{equation}
is imposed then for the $\gamma_\ell(r,r')$ functions we have only the potential-independent restrictions of
\begin{equation}
L_{r0} \gamma_\ell(r,r')=L_{r'0} \gamma_\ell(r,r'),\quad \gamma_\ell(r,0)=\gamma_\ell(0,r')=0,\quad\forall\ell.
\end{equation}
and the information is only stored in the $c_\ell$ expansion coefficients.

\section{Cox--Thompson (CT) method}

In the framework of the method proposed by Cox and Thompson
\cite{CT1} one takes the following separable form for the
$\gamma_\ell(r,r')$ functions
\begin{equation}\label{CTg}
\gamma_\ell(r,r')=u_\ell(\min(r,r'))v_\ell(\max(r,r')),
\end{equation}
which is also the Green's function of the $q\equiv0$ radial Schr\"odinger
equation for the $\ell$th partial wave:
\begin{equation}
\left[\frac{d^2}{dr^2}+1-\frac{\ell(\ell+1)}{r^2}\right]\gamma_\ell(r,r')=\delta(r-r');
\end{equation}
and the summation in $g(r,r')$ runs only over a finite set $S$ of $\ell$'s:
\begin{equation}\label{CTg1}
g(r,r')=\sum_{\ell\in S} c_\ell u_\ell(\min(r,r'))v_\ell(\max(r,r')),
\end{equation}
containing the Riccati-Bessel functions, connected to the Bessel and Neumann
functions by
\begin{equation}
u_\ell(r)=\sqrt{\frac{\pi r}{2}}J_{\ell+\frac{1}{2}}(r),\qquad v_\ell(r)=\sqrt{\frac{\pi r}{2}}Y_{\ell+\frac{1}{2}}(r).
\end{equation}

For solving the GLM-type equation (\ref{GLM}) we use the separable ansatz
\begin{equation}
K(r,r')=\sum_{L\in T}A_L(r)u_L(r')
\end{equation}
with a finite set $T$ of "shifted angular momenta" satisfying
$S\cap T=\emptyset$ and $|S|=|T|$. The use of such an ansatz can
be motivated by the following result ascertained from \cite{CT2}.
\begin{proposition}
If $r'^{-1/2}K(r,r')=O(1)$, $r'\to0$ is imposed on $K(r,r')$ and the $L$
numbers are restricted to $L>-0.5$ then
\begin{equation}
c_\ell=\frac{\prod_{L\in T}(\ell(\ell+1)-L(L+1))}{\prod_{\ell'\in S, \ell'\neq\ell}(\ell(\ell+1)-\ell'(\ell'+1))}\quad\Leftrightarrow\quad K(r,r')=\sum_{L\in T}A_L(r)u_L(r').
\end{equation}
Where $\{c_\ell\}\leftrightarrow\{L\}$ is a one-to-one mapping.
\end{proposition}

Also, this form provides a reasonably easy way to solve the GLM-type integral
equation. In fact it makes a system of algebraic equations instead of the
integral equation, namely
\begin{equation}\label{GLMalg}
\sum_{L\in T}A_L(r)\frac{u_L(r)v'_\ell(r)-u'_L(r)v_\ell(r)}{\ell(\ell+1)-L(L+1)}=v_\ell(r),\qquad \ell\in S.
\end{equation}

The parameters of the set $T$ can be obtained from the phase shifts
$\{\delta_\ell\}_{\ell\in S}$ through the transformation equation (\ref{PL})
which in terms of the $L$'s takes the form
\begin{equation}\label{PL1}
\varphi_\ell(r)=u_\ell(r)-\sum_{L\in T}A_L(r)\frac{u_L(r)u'_\ell(r)-u'_L(r)u_\ell(r)}{\ell(\ell+1)-L(L+1)}
\end{equation}
Taking both equation (\ref{PL1}) and (\ref{GLMalg}) for large $r$'s, i.e. at $r\to\infty$ we get the system of nonlinear equations \cite{AHS2003,MSA2006,PHA2008}
connecting $\{\delta_\ell\}_{\ell\in S}$ to $T$:
\begin{equation}\label{NL}
e^{2i\delta_\ell}=\frac{1+i\mathcal{K}_\ell^+}{1-i\mathcal{K}_\ell^-},\qquad \ell\in S,
\end{equation}
with
\begin{equation}\mathcal{K}_\ell^{\pm}=\sum_{L\in T}\sum_{\ell'\in S}[M_{\sin}]_{\ell L}[M_{\cos}^{-1}]_{L\ell'}e^{\pm i(\ell-\ell')\pi/2}\qquad \ell\in S,
\end{equation}
\begin{equation}
\left\{\begin{array}{ll}
M_{\sin}\\
M_{\cos}
\end{array}\right\}_{\ell L} = \frac{1}{L(L+1)-\ell(\ell+1)}\left\{ \begin{array}{ll}
\sin\left((\ell-L)\pi/2\right)\\
\cos\left((\ell-L)\pi/2\right)
\end{array} \right\},\quad \ell\in S,\,L\in T.
\end{equation}
After some investigation of this nonlinear system of equations one can conclude that for a given set of phase shifts it yields an infinity of solutions, each giving rise to a potential. It was our result in \cite{PAJMP} that one can select a physical potential (perhaps uniquely) with the aid of a consistency check. This check is first revisited in Sec. 4 and then it is applied for the special cases of one  and two dimensions, $|S|=|T|=1,\,2$.

\section{Some new results concerning the CT potential}

Before continuing with the consistency check it is worthwhile to further study the CT method particularly concerning the potential one can obtain by applying it.
We address the case when only a finite number of input phase shifts are used to construct the potential.

\subsection{Asymptotics}

First, we show that the potential is generally not compactly supported nor is of long-range. To see this define the functions $\{A^a_L(r)\}_{L\in T}$ by the limit
\begin{equation}
A_L(r)=A^a_L(r)+o(1),\qquad r\to\infty,\quad L\in T.
\end{equation}
We infer from \cite{PHA2008} that the asymptotic coefficients $A^a_L(r)$ assume the form
\begin{equation}\label{trigo}
A_L^a(r)=a_L\cos r+b_L\sin r
\end{equation}
where $a_L$'s and $b_L$'s are constant depending on all the elements of both $S$ and $T$:
\begin{equation}\label{aL}
\sum_{L\in
T}a_{L}\frac{\cos(\frac{\pi}{2}(\ell-L))}{L(L+1)-\ell(\ell+1)}=\cos\left(\ell\frac{\pi}{2}\right),\qquad
\ell\in S,
\end{equation}
\begin{equation}\label{bL}
\sum_{L\in
T}b_{L}\frac{\cos(\frac{\pi}{2}(\ell-L))}{L(L+1)-\ell(\ell+1)}=\sin\left(\ell\frac{\pi}{2}\right),\qquad
\ell\in S.
\end{equation}
This form in turn implies that
\begin{equation}\label{Kass}
K(r,r)=\alpha\sin(2r)+\beta\cos(2r)+\gamma+o(1),\qquad r\to\infty
\end{equation}
and thus
\begin{equation}
q(r)=4\frac{\beta\sin(2r)-\alpha\cos(2r)}{r^2}+o(r^{-2}),\qquad r\to\infty
\end{equation}
which means that the potential generally falls of like an inverse power of two.

One can give a necessary condition for the potential to decrease more rapidly then $O(r^{-2})$. One only needs 
\begin{equation}\label{sumr1}
\alpha=\beta=0.
\end{equation}
The quantites $\alpha$ and $\beta$ are given by
\begin{align}
\alpha&=\frac{1}{2}\sum_{L\in T} \left(a_L\cos L\frac{\pi}{2}-b_L\sin L\frac{\pi}{2}\right)\\
\beta&=-\frac{1}{2}\sum_{L\in T} \left(a_L\sin L\frac{\pi}{2}+b_L\cos L\frac{\pi}{2}\right).
\end{align}
Alternatively, $K(r,r)$ can be written as a series in $\ell$ (this formula does not hold for $K(r,r')$ with $r\neq r'$),
\begin{equation}
K(r,r)=\sum_{\ell\in S}c_\ell \varphi_{\ell}(r) v_\ell(r),
\end{equation}
which can readily be seen from the GLM equation (\ref{GLM}) taken at $r=r'$ and the CT formula (\ref{CTg1}) substituted for $g(r,r')$:
\begin{align}
K(r,r)&=\sum_{\ell\in S}c_\ell u_\ell(r)v_\ell(r)-\int_0^r d\rho\rho^{-2}K(r,\rho)\sum_{\ell\in S}c_\ell u_\ell(\rho)v_\ell(r)\\
&=\sum_{\ell\in S}c_\ell v_\ell(r)\left[u_\ell(r)-\int_0^r K(r,\rho)u_\ell(\rho)\rho^{-2}d\rho\right],
\end{align}
where the formula (\ref{PL}) for $\varphi_\ell(r)$ has appeared. This allows for the alternative conditions
\begin{equation}\label{sumr2}
\sum_{\ell\in S}(-1)^\ell c_\ell B_\ell \cos\delta_\ell=0,\quad\text{and}\quad\sum_{\ell\in S}(-1)^\ell c_\ell B_\ell \sin\delta_\ell=0
\end{equation}
involving the expansion coefficients, the input phase shifts and the normalization constants of the partial wave functions.

These new results may serve as useful tools to check numerical results or incorporated into a solution method they might provide a way to control the undesirable oscillations of the inverse potential.

At this point we shortly discuss the applicability of the CT method. The inverse potential is finite at the origin starting with a zero derivative (see e.g. \cite{PSA2009}) and possesses also a finite first moment $\int_0^\infty r q(r) dr<\infty$. At fixed scattering energy the CT procedure is thus a particularly successful method for (re)constructing inverse potentials. It was used to recover interaction potentials (see e.g. \cite{AHS2003,MSA2006,PHA2008}) for various physical system. Furthermore it is also possible to generalize the method to recover potentials having Coulomb tail e.g. by replacing the Riccati-Bessel function by Coulomb wave functions \cite{PSA2009,HSPA2010}.

\subsection{CT potentials regarding statistics of colliding particles}

In real-life scattering experiments one encounters cases when the colliding particles are zero spin bosons. Then it is well-known that only the even numbered partial waves contribute to the scattering amplitude because of the symmetry of the wave function. In such cases with the CT method it is possible to use only the experimentally available phase shift data corresponding to even partial waves. It is then reasonable to ask for the value of phase shifts provided by the CT potential for the odd partial waves. It turns out that these odd phase shifts are zero. 

To see this let us derive the phase shifts of the CT potential which can be obtatined from Eq. (\ref{PL1}) taken at $r\to\infty$:
\begin{equation}
\varphi_j(r)=\sin(r-j\pi/2)-\sum_{L\in T}A^a_L(r)\frac{\sin\left((j-L)\pi/2\right)}{j(j+1)-L(L+1)}+o(1),\quad r\to\infty.
\end{equation}
Now the trigonometric form (\ref{trigo}) is substituted for $A_L^a(r)$. There are two special cases that we consider:\\
i) the set $S$ of input angular momenta consists of only even numbers,
\\
ii) $S$ consists of odd numbers.\\
In the first case we have $b_L=0$ $\forall L\in T$ while in the second $a_L=0$ $\forall L\in T$. This is deductible from Eqs. (\ref{aL}) and (\ref{bL}) which assume the forms
\begin{eqnarray}
\sum_{L\in T}\left\{\begin{array}{ll}
a_L\\
b_L
\end{array}\right\}\frac{\cos\left(L\frac{\pi}{2}\right)}{L(L+1)-\ell(\ell+1)} = \left\{\begin{array}{ll}
1\\
0
\end{array} \right\},\qquad \ell\in S\\
\sum_{L\in T}\left\{\begin{array}{ll}
a_L\\
b_L
\end{array}\right\}\frac{\sin\left(L\frac{\pi}{2}\right)}{L(L+1)-\ell(\ell+1)} = \left\{\begin{array}{ll}
0\\
1
\end{array} \right\},\qquad \ell\in S
\end{eqnarray}
for the cases i) and ii) respectively. One can assume $\cos\left(L\frac{\pi}{2}\right)\neq0$ and $\sin\left(L\frac{\pi}{2}\right)\neq0$, since the matrix $M$ with elements $M_{\ell L}=(L(L+1)-\ell(\ell+1))^{-1}$ is invertible (it is a Cauchy matrix) and cannot be singular unless $T\cap S\neq\emptyset$ which is excluded by assumption. We thus have $b_L=0$ $\forall L\in T$ for i) and $a_L=0$ $\forall L\in T$ for ii). Now 
\begin{equation}
\sin(r-j\pi/2)=
\begin{cases}
(-1)^j\sin r, & \text{$j$ even}
\\
(-1)^{j+1}\cos r, &\text{$j$ odd}
\end{cases}
\end{equation}
which implies
\begin{equation}
\varphi_j(r)=B_j\sin(r-j\pi/2)+o(1)\qquad\text{for i) and odd $j$, or for ii) and even $j$}.
\end{equation}
In other words the CT phase shifts of the opposite parities are exactly zero. Notice that the CT method allows the construction of potentials which are transparent for half the partial waves (being even or odd in parity).

Applications suggest that if dealing with input partial wave data of one parity the performance of the CT method with the same number of input diminishes compared to the case when data with both parity is employed. Therefore our sum rules derived in the previous subsection can be extremely useful to improve performance by supressing oscillations of the potential. Also note that one of our sum rules simplify (due to $B_\ell\cos\delta_\ell=1,\,\ell\in S$ in the even case and $B_\ell\sin\delta_\ell=1,\,\ell\in S$ in the odd case \cite{PHA2008proc}) to
\begin{equation}
\sum_{\ell\in S} c_\ell=0
\end{equation}
while the other becomes $\sum_{\ell\in S} c_\ell\tan\delta_\ell=0$ and $\sum_{\ell\in S} c_\ell\cot\delta_\ell=0$ for the even and the odd case, respectively.

\section{Consistency check}
\subsection{General condition}

The kernel $g(r,r')$ only makes sense if the integral equation is
uniquely solvable with it. As Eq. (\ref{GLM}) can be viewed as a
Fredholm type integral equation of the second kind for fixed $r$,
viz. if $rr'\kappa(r,r')=K(r,r')$ and $rr'\gamma(r,r')=g(r,r')$ we
have
\begin{equation}
\kappa(r,r')=\gamma(r,r')-\int_0^r\gamma(r',\rho)\kappa(r,\rho)d\rho.
\end{equation}
According to Fredholm's alternative the Fredholm determinant thereof must be
nonzero for all fixed $r>0$. When using the CT ansatz the Fredholm determinant
of the integral equation becomes the determinant of the system of the algebraic
equations (\ref{GLMalg}),
\begin{equation}
D(r)=\det\left\{\left[\frac{u_L(r)v'_\ell(r)-u'_L(r)v_\ell(r)}{\ell(\ell+1)-L(L+1)}\right]_{\ell L}\right\}.
\end{equation}
It is easy to see that for $X\in\Omega=\{x\,:\,D(x)=0,\, x\in\mathbb{R}^+\}$
one gets
\begin{equation}
\lim_{x\to X}\int_0^xtq(t)dt=\pm\infty,
\end{equation}
thus the potential is not in $L_{1,1}=\{q\,:\,\int_0^\infty t|q(t)|dt<\infty\}$.
While for $\Omega=\emptyset$ we have \cite{CT1}
\begin{equation}
\int_0^\infty tq(t)dt=\sum_{L\in T}\frac{\prod_{\ell\in S}(L-\ell)}{\prod_{L\neq L'\in T}(L-L')}<\infty.
\end{equation}

One can conclude the following \cite{PAJMP}.
\begin{theorem}
If and only if $D(r)\neq0$ on $r>0$ we get a unique solution of the GLM-type
integral equation in $C^2(\mathbb{R}^+\times\mathbb{R}^+)$ and from that an
inverse potential in $L_{1,1}=\{q\,:\,\int_0^\infty t|q(t)|dt<\infty\}$.
\end{theorem}

From \cite{RammAA} we know that $D(r)\neq0$ on $r>0$ is not the case for
arbitrary choice of $S$ and $T$: e.g. there it was shown, that if $S=\{0\}$ and
$L=\{2\}$ then $D(r)=0$ at some $r>0$.

With the help of Theorem 4.1 we can convince ourselves that a particular CT inverse potential obtained numerically from arbitrary data is integrable or not. If it is so then that potential will generate the input phase shifts.

Also, using the theorem it is possible to determine the admissible set of $L$ numbers for given $\ell$'s. In the one-$\ell$ case we get a straightforward admissible set of $L$'s, however in higher dimensions the formulation becomes extremely involved.

\subsection{One dimensional case}
In this case -- $S=\{\ell\}$, $|S|=1$ -- the function
$W(u_L,v_\ell)(r)\equiv u_L(r)v'_\ell(r)-u'_L(r)v_\ell(r)$ must be examined
carefully. This was performed in \cite{PAJMP} and thus it shall not be
discussed here in detail. We only give the key idea of the proof: the Wronskian
\begin{equation}
\frac{u_L(r)v'_\ell(r)-u'_L(r)v_\ell(r)}{\ell(\ell+1)-L(L+1)}
\end{equation}
is nonzero on $\mathbb{R}^+$ if and only if the constituent Bessel functions
($J_{L+1/2}(x)$ and $Y_{\ell+1/2}(x)$) are interlaced. This is deductible from
the observation
\begin{equation}
D(r)=\frac{u_L(r)v'_\ell(r)-u'_L(r)v_\ell(r)}{\ell(\ell+1)-L(L+1)}=\int_0^ru_L(\rho)v_\ell(\rho)\rho^{-2}d\rho.
\end{equation}

We could prove the next theorem of broader interest \cite{PAJMAA}
\begin{theorem}
The positive zeros of the Bessel functions $J_\nu(x)$, $J'_\nu(x)$, $Y_\nu(x)$,
$Y'_\nu(x)$, $J_{\nu+\varepsilon}(x)$, $Y_{\nu+\varepsilon}(x)$ for nonnegative
orders, $\nu\geq0$ are interlaced according to the inequalities
\begin{equation}
\nu\leq j'_{\nu,1}<y_{\nu,1}<y_{\nu+\varepsilon,1}<y'_{\nu,1}<j_{\nu,1}<j_{\nu+\varepsilon,1}<j'_{\nu,s+1}<\ldots
\end{equation}
if and only if $0<\varepsilon\leq1$ (otherwise $y_{\nu+\varepsilon,s}>j_{\nu,s}$ for some $s$
and $y_{\nu,s'+1}>j_{\nu+\varepsilon,s'}$ for some $s'$).
\end{theorem}
The following gives the admissible set $|\ell-L|\leq1$.
\begin{theorem}
 $W(u_L,v_\ell)(r)$ has no roots on $r\in\mathbb{R}^+$, that is at $N=1$ the
 GLM-type equation is uniquely solvable for the CT method with $S=\{\ell\}$
 and $T=\{L\}$ if and only if $|L-\ell|\leq1$. $\ell\in(-0.5,\infty)$,
 $L\in(-0.5,\infty)$ is supposed.
\end{theorem}

This result allows us to choose {\it uniquely} from the solutions of the system of equations (\ref{NL}) as the solution for one input phase shift is
\begin{equation}\label{CT1s}
L=\ell-\frac{2}{\pi}\delta_\ell+2k,\qquad k\in\mathbb{Z},
\end{equation}
which for $\delta_\ell\in\left[-\frac{\pi}{2},\frac{\pi}{2}\right]$ in conjunction with the previous theorem yields
\begin{equation}\label{CT1s2}
L=\ell-\frac{2}{\pi}\delta_\ell.
\end{equation}

At low energies it may happen that only one partial wave contributes mostly to the scattering amplitude (e.g. for some partial wave resonances). It is worthwhile to look at the phase shifts yielded by the CT inverse potential at the one-phase-shift level to get a sense of the quality of the inversion procedure.

Deductible from Eq. (\ref{PL1}) is the following formula
\begin{equation}\label{1shph}
\tan\delta_\ell= 
\begin{cases} 0, & \text{$\ell$ odd}
\\
\frac{L(L+1)}{L(L+1)-\ell(\ell+1)}\tan\delta_0, &\text{$\ell$ even}
\end{cases}
\end{equation}
which specifies the phases of the CT potential for $S=\{0\}$.

If e.g. the phase shift is restricted to describe an attractive potential (i. e. $\delta_0<0$) the bound
\begin{equation}
\tan\delta_\ell\leq
\begin{cases} 0, & \text{$\ell$ odd}
\\
\frac{4}{15\ell^2}\tan\delta_0, &\text{$\ell$ even}
\end{cases}
\end{equation}
can be found using Eq. (\ref{1shph}). This result assures the proper reproduction of the phase shifts by the CT potential.

To illustrate these results  Fig. 1 shows synthetic test potentials
corresponding to $\ell=0$, $\delta_0=0.2\pi$ obtained by the CT method. In addition
to the $L_{1,1}$ potential an inconsistent one is also shown where $k$ in Eq.
(\ref{CT1s}) is chosen to be other than zero. As indicated before we get a
non-integrable potential.

\begin{figure}[h!]
  \label{fig1}
  \caption{Potentials obtained by the CT method corresponding to $\ell=0$,
  $\delta_0=0.2\pi$ with $L=-0.4$ (solid line) and $L=1.6$ (dashed line).}
  \centering
    \includegraphics[width=0.8\textwidth]{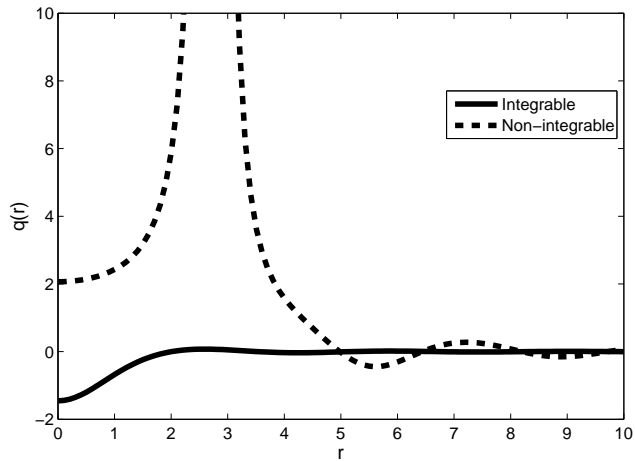}
\end{figure}

\subsection{Two dimensional case}
Already in this case the Fredholm determinant becomes complicated, which is also
apparent from it's integral representation
\begin{equation}
D(r)=\int_0^r\int_0^r u_{L_1}(\rho)u_{L_2}(\rho')(v_{\ell_1}(\rho)v_{\ell_2}(\rho')-v_{\ell_1}(\rho')v_{\ell_2}(\rho))\rho^{-2}\rho'^{-2}d\rho d\rho'.
\end{equation}
Therefore instead of the analytical treatment we determine the admissible set of
$L$'s numerically for some choices of $\ell$'s.

Our numerical method was to check the determinant at the points of a fine lattice
on the $L_1$--$L_2$ quarter plane of $\mathbb{R}^+\times\mathbb{R}^+$ whether it
has any zeros on $(0,\Lambda)$, where $\Lambda$ is a large number chosen to be
great enough for
\begin{equation}
D(r>\Lambda)=\text{const.}+\varepsilon(r),\qquad |\varepsilon(r)|<\varepsilon
\end{equation}
with a very small $\varepsilon$. This can be done since every $D(r)$ in any
dimensions $|S|<\infty$ has only a finite number of zeros, since the constituent
Wronskians all tend to constants at large $r$ distances, viz.
\begin{equation}
W(r)=u_L(r)v'_\ell(r)-u'_L(r)v_\ell(r)=\cos\left[(\ell-L)\frac{\pi}{2}\right]+O\left(\frac{1}{r}\right),\quad r\to\infty.
\end{equation}

On Fig. 2 the admissible sets of $\{L_1,L_2\}$ pairs are depicted for the
particular choices  $S_1=\{1,3\}$ and $S_2=\{1,2\}$.

\begin{figure}[h!]
  \label{fig2}
  \caption{The admissible sets of the $T$ elements for (a) $S_1=\{1,3\}$ and
  (b) $S_2=\{1,2\}$ denoted by blank areas.}
  \centering
    \subfloat[]{\includegraphics[width=0.5\textwidth]{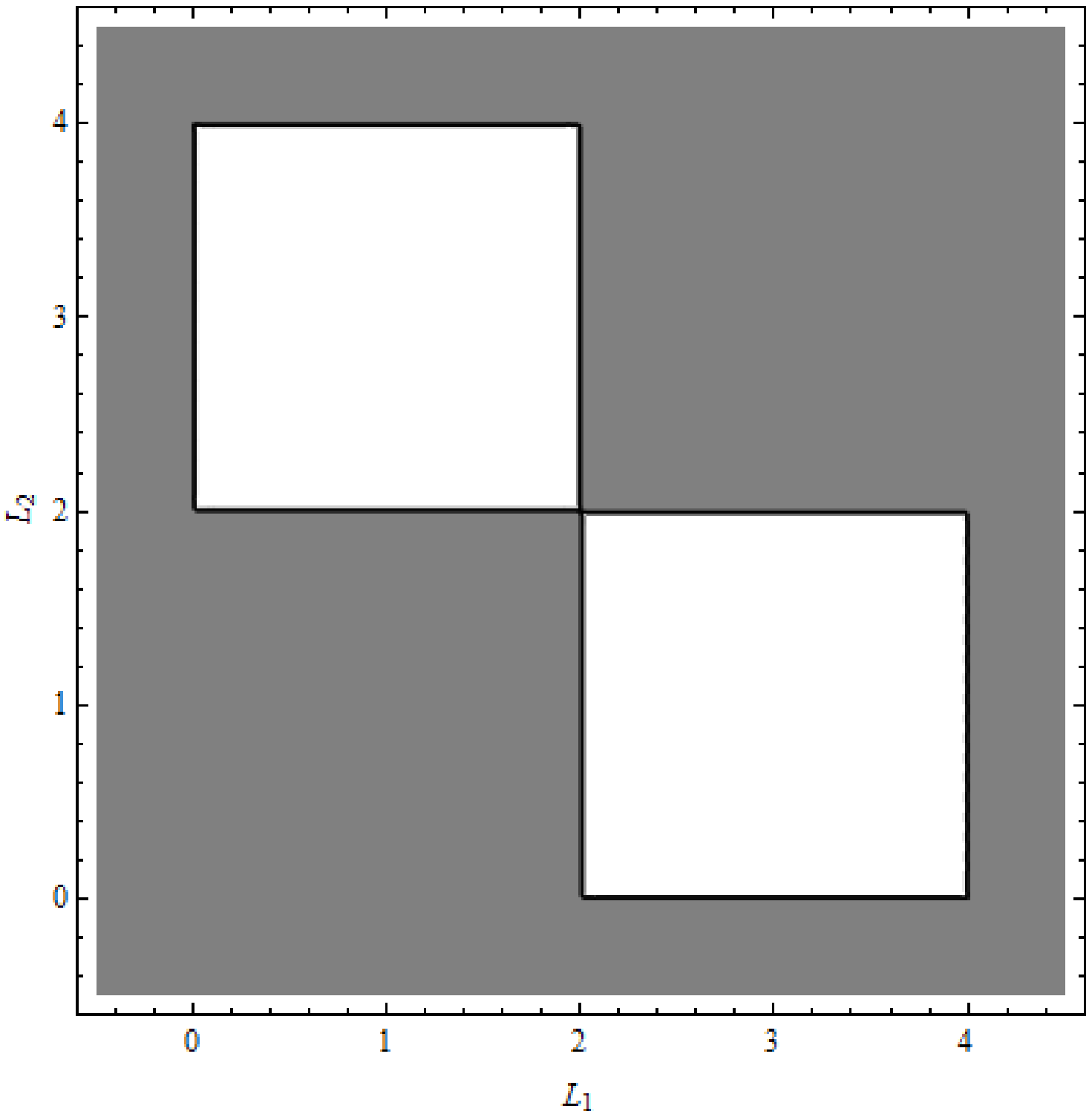}}
    \subfloat[]{\includegraphics[width=0.5\textwidth]{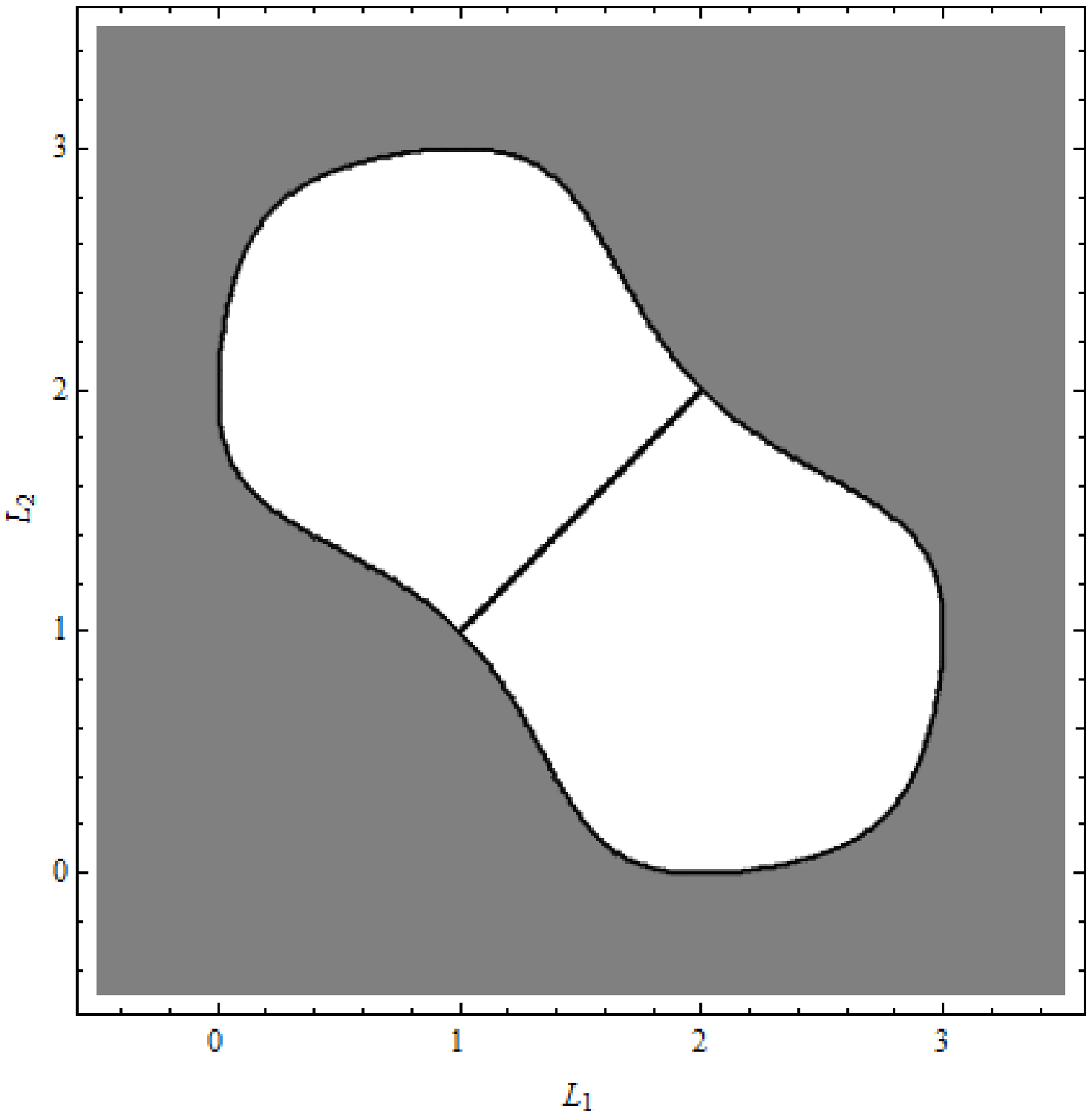}}
\end{figure}

In the next example (Fig. 3) we calculated some possible $\{L_1,L_2\}$ pairs for a given $\{\delta_{\ell_1},\delta_{\ell_2}\}$. We note that only one of them is inside the permitted domain and could only find a single solution of the system of non-linear equation that is permitted by the consistency condition. Again the $L_{1,1}$ and an inconsistent potential is shown.

\begin{figure}[h!]
  \label{fig3}
  \caption{Potentials obtained by the CT method corresponding to $S=\{0,1\}$,
  with phase shifts calculated from a Woods-Saxon potential.
  $T=\{-0.3056,0.9295\}$ (solid line) and $T=\{1.0650,1.7016\}$ (dashed line).
  Also, the original potential, $q(r)=-\left[1+e^{2.5\cdot(r-1)}\right]^{-1}$, yielding the phase shifts ($\delta_0=0.4389$, $\delta_1=0.1246$) is depicted (dotted line).}
  \centering
    \includegraphics[width=0.9\textwidth]{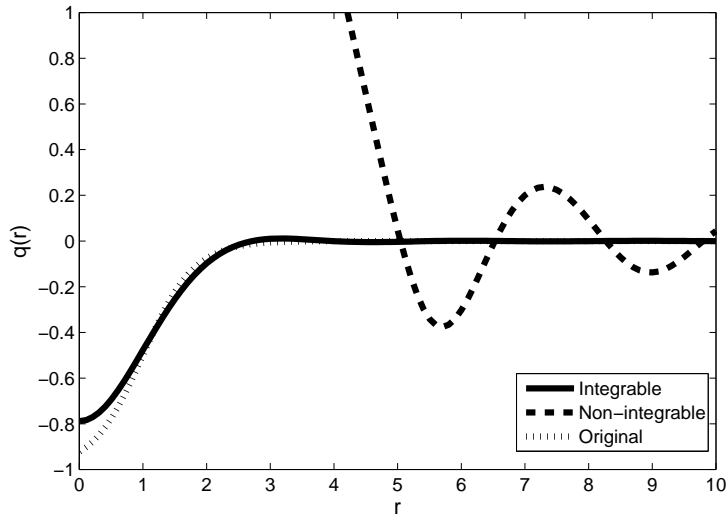}
\end{figure}

\section{Conclusion}
After recalling some crucial results of the inverse scattering problem for the Schr\"odinger equation at fixed scattering energy we reviewed a particular constructive scheme, the Cox-Thompson (CT) method for recovering scattering potentials responsible for a finite set of given phase shifts.

We have analyzed the asymptotic properties of the CT potentials and from that we have obtained a sum rules for the expansion coefficients (Eq. (\ref{sumr1},\ref{sumr2})). Then we considered CT potentials corresponding to specific quantum statistical characters of the colliding pattern. If the scattered particles are zero spin bosons (e.g. oxygen atoms or carbon nuclei) the CT potentials are constructed from even phase shifts being accessible for measurements. We have shown that in this case the CT potentials in addition to reproducing the input even phase shifts give exactly zero phases for the odd partial waves.

Then we turned to the consistency check. It has been shown that for the potential to be integrable (or more precisely, to belong to the function class $L_{1,1}$) a condition must be fulfilled for the intermediate quantities $L$ of shifted angular momenta of the CT method. We discussed the case when only one input phase shift is used in which circumstance we have explicit uniqueness. The two-phase-shift case has also been discussed. The condition for the $L$ numbers to produce an integrable potentials has been obtained through a numerical calculation for two particular cases and the results have been illustrated in Fig.3.

\end{document}